# On Parallel or Distributed Asynchronous Iterations with Unbounded Delays and Possible Out of Order Messages or Flexible Communication for Convex Optimization Problems and Machine Learning

Didier El-Baz
*LAAS-CNRS*
*Université de Toulouse, CNRS*
Toulouse, France
elbaz@laas.fr

*Abstract*— We describe several features of parallel or distributed asynchronous iterative algorithms such as unbounded delays, possible out of order messages or flexible communication. We concentrate on the concept of macro-iteration sequence which was introduced in order to study the convergence or termination of asynchronous iterations. A survey of asynchronous iterations for convex optimization problems is also presented. Finally, a new result of convergence for parallel or distributed asynchronous iterative algorithms with flexible communication for convex optimization problems and machine learning is proposed.

I. INTRODUCTION

Iterative algorithms are successive approximation methods which are particularly efficient, when combined with parallel or distributed schemes of computation, to solve several classes of large-scale numerical simulation problems or optimization problems. Amongst these methods, parallel or distributed asynchronous iterative algorithms, whereby iterations are carried out in parallel or distributed way in arbitrary order and without synchronization, have been studied for a long time on a theoretical and practical manner, e.g. see: [1], [2], [3] and [4].

Today, asynchronous block relaxation methods are very popular as smoothers for multigrid methods and in combination with subdomain preconditioners for Krylov methods [5].

Asynchronous iterative algorithms have also been applied with success to the solution of convex optimization problems, e.g., see: [6],[7], [8], [9] and [10]. Distributed asynchronous iterations which aim at solving a global problem via local data have received considerable attention in the context of convex optimization [11].

When it comes to the analysis of distributed asynchronous iterative algorithms, the major difficulties come from the lack of global clock and absence of synchronization. In particular, the study of convergence, the detection of the convergence and the analysis of the convergence rate of asynchronous iterative algorithms are complex issues.

Research work funded by CIMI-ANR-11-LABX-0040 - LABX-2011.

In this paper we survey important concepts and results in the domain of parallel or distributed asynchronous iterative algorithms for convex optimization problems. In particular, we consider the concepts of unbounded delays, out of order messages, macro-iteration sequence and flexible communication. We analyze the advantages of parallel or distributed asynchronous iterations. Finally, we present a new convergence result for parallel or distributed asynchronous iterations with flexible communication with application to convex optimization and machine learning applications.

Section II deals with related works. Unbounded delays and possible out of order messages are studied. Macro-iteration sequence is analyzed in Section III. Asynchronous iterations with flexible communication are presented in section IV. A new convergence result for parallel or distributed asynchronous iterations with flexible communication for convex optimization and machine learning applications is proposed in section V. Conclusions and future work are presented in section VI.

II. RELATED WORKS

In pioneering works Chazan and Miranker [12], Rosenfeld [13] and Miellou [14] have proposed and studied chaotic relaxation methods for the solution of linear and nonlinear fixed-point problems, respectively.

The mathematical model of chaotic relaxation features bounded delays in order to represent the nondeterministic behavior of the proposed parallel iterative methods, whereby computations are carried out in arbitrary order and without synchronization.

Baudet introduced the general model of asynchronous iterations with unbounded delays in [1]. Asynchronous iterative methods are defined as follows.

*Definition 1:* Let $F$ be an operator from $\mathbb{R}^n$ to $\mathbb{R}^n$, for all $i \in \{1, \dots, n\}$, let the functions $l_i: N \to N$, an asynchronous iteration $(F, x(0), S, \mathcal{L})$ associated to the operator $F$ and starting with the given iterate vector $x(0)$, is a sequence $\{x(j)\}, j = 0, 1, \dots$, of vectors of $\mathbb{R}^n$ defined recursively by:

$$x_i(j) = \begin{cases} F_i\big(x_1(l_1(j)), \dots, x_n(l_n(j))\big) & if\ i \in S_j, \\ x_i(j-1) & if\ i \notin S_j, \end{cases} \quad (1)$$

where $\mathcal{S} = \{\mathcal{S}_j | j = 1, 2, ...\}$ is a sequence of nonempty subsets of $\{1, ..., n\}$ and $\mathcal{L} = \{(l_1(j), ..., l_n(j))\}$, $j = 1, 2, ...$ is a sequence of tuples in $\mathbb{N}^n$. In addition, $\mathcal{S}$ and $\mathcal{L}$ are subject to the following conditions for all $i \in \{1, ..., n\}$:

  a) $l_i(j) \leq j - 1, j = 1, 2, ...$;
  b) $l_i: \mathbb{N} \to \mathbb{N}$, is such that $\lim_{j \to \infty} l_i(j) \to +\infty$;
  c) $i$ occurs infinitely often in the sets $\mathcal{S}_j, j = 1, 2, ...$

The above mathematical model details the updating phases of asynchronous iterations. The vector $x(0)$ corresponds to an initial approximation of the solution of the problem. The components of the iterate vector $x$ are updated in parallel or in a distributed way. Since the global behavior of asynchronous iterations is non-deterministic, the set $\mathcal{S}$ accounts for all possible steering policies, i.e., for the choice of the set of components of the iterate vector that are updated (or relaxed) at each iteration. This feature contributes to make asynchronous iterations a general model of parallel or distributed iterative algorithms. The set $\mathcal{L}$ accounts for the labels of the iterations which are used at each updating phase.

Condition *a)* accounts for the obvious fact that the value used at each updating phase are produced at previous updating phases, i.e., previous iterations.

It follows from assumptions *a)* and *b)* that labels $l_i(j)$ correspond to past updates, i.e. delayed components of the iterate vector. We note that the introduction of delays in the mathematical model of asynchronous iterations does not imply that parallel or distributed asynchronous iterative methods are not efficient. Delays are a convenient way to represent the lack of synchronicity and nondeterministic behavior of asynchronous iterations. The use of unbounded delay accounts for representing a general model of parallel or distributed iterative methods. In particular, condition *b)* models the unpredictable behavior of asynchronous iterative algorithms that do not feature synchronization mechanisms. It also accounts for unbounded delays and possible out of order messages or updates. This last point is particularly important in practice when considering distributed systems or distributed memory architectures whereby data are exchanged via message passing.

Condition *c)* implies that no component of the iterate vector is abandoned forever during the global updating process.

We highlight that one can hardly prove that asynchronous iterative algorithms converge without conditions *b)* and *c)*.

Figure 1 displays an example of parallel or distributed asynchronous iterative algorithm. For facility of presentation, we consider a case with two processors denoted by $P_1$ and $P_2$, respectively. Figure 1 shows the sequence of updating phases as time progresses. Each rectangle corresponds to an updating phase that is labelled by an iteration number. For easiness of understanding, we can think about a fixed-point operator $F$ from $\mathbb{R}^2$ to $\mathbb{R}^2$, where component $x_1$ is updated by processor $P_1$ and component $x_2$ is updated by processor $P_2$.

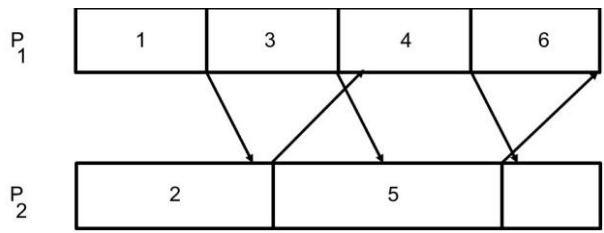

Fig. 1 Example of parallel or distributed asynchronous iterative algorithm.

In Figure 1, data exchange is represented. It can be performed via message passing or writing in a shared memory. The value of the components of the iterate vector that are labelled by an iteration number are communicated at the end of each updating phase. Arrows represent operations of communication between processors. For example, the first arrow from processor $P_1$ to processor $P_2$ corresponds to the communication of $x_1(1)$. The second arrow from processor $P_2$ to processor $P_1$ corresponds to the communication of $x_2(2)$.

Each processor goes at its own pace. Processors perform updating phases according to their computing load. We note that computations are covered by communication. As a consequence, there are no idle time resulting from communication. Similarly, there is no waiting due to synchronization.

The general model of asynchronous iterations features unbounded delays. In [1], Baudet gives the following simple example where delays are unbounded and condition *b)* is satisfied. Let processor $P_1$ always update component $x_1$ in one unit of time and assume that it takes $k$ units of times for processor $P_2$ to perform the $k - th$ updating phase of component $x_2$. Then, a simple calculation shows that the delay in updating component $x_2$ grows as $\sqrt{j}$ and we have:

$$\lim_{j \to \infty} l_2(j) = \lim_{j \to \infty} j - \sqrt{j} \to +\infty.$$

Chazan, Miranker and Miellou have considered bounded delays in the specific case of chaotic relaxation (see [12] and [14]). They have proposed a mathematical model of chaotic iterations with functions $l_i: \mathbb{N} \to \mathbb{N}$, $i \in \{1, ..., n\}$ relative to delayed labels such that $l_i(j) = j - d_i(j)$, for all $i \in \{1, ..., n\}$. These functions satisfy the following condition:

  *d) there exists a natural number $b > 0$ and a function* $b(j): \mathbb{N} \to \mathbb{N}$, *such that*

  1) $b(j) \leq \min\{b, j\}$,
  2) $j - b(j)$ *is a monotone increasing function of* $j$,
  3) $\forall j \in \mathbb{N}, \forall i \in \{1, ..., n\}, 0 \leq d_i(j) < b(j)$.

In the case of chaotic relaxation methods, the natural number $b$ in assumption *d)* represents the bound on the delay to access the previous updates that are used in computations.

Later, Miellou studied asynchronous iterations with unbounded delays, e.g., see [15]. Miellou published important theoretical contributions in the domain; most of them are related to numerical simulation. At that time there, was a bloom of scientific publications of the "French school" on parallel or distributed asynchronous iterative algorithms (in particular, see [2] and [4]) and some experts, such as Professor Yousef Saad, have pointed that the French research works were "visionary" and "in advance of several decades" (see [16]).

From the late seventies to the beginning of the new millennium, the motivation of researchers of the French school came from the conviction that future supercomputers and High-Performance Computing systems will be massively parallel machines with hundred thousand processors and that synchronization was the major performance-limiting factor, in this context. Capacity of integration of microprocessors was always increasing according to Moore's law and clock frequency was clearly a limiting factor with regards to energy consumption. We note that we have implemented asynchronous iterative algorithms on supercomputers such as the Cray T3E and IBM SP4 for convex optimization problems and numerical simulation problems in [10] and [26], respectively.

Another motivation of our research work on distributed asynchronous iterations was that distributed systems will follow the same trend. As a matter of fact, in the context of packet-switched network, the first routing algorithm to be implemented on the Arpanet in 1969 was a distributed asynchronous Bellman-Ford algorithm (see [11], pp. 479-480). We recall that Arpanet was the precursor of the Internet. The asynchronous Bellman-Ford algorithm was also used in other data communication networks, see [17]. The concepts of modular robotics and distributed autonomous robotic systems also supported new trends in favor of massively distributed systems. These concepts have applications in fault-tolerant, resilient, perdurable, and reconfigurable robotics systems whose duration and profitability are increased, as well as in the concept of programmable matter which aims at featuring ten thousand tiny modules [18], [19], [20] and [21].

The main advantages of parallel or distributed asynchronous iterative algorithms are:
- to get rid of waiting time resulting from synchronization;
- to recover communication by computation;
- to cope naturally with load unbalancing; this property is particularly important in the case of massive parallelism or distribution.

These properties generally contribute to improve efficiency and scalability of parallel or distributed iterative methods. Aside, the lack of synchronization leads to some fault-tolerance, e.g., transient faults in data exchange are covered by the arrival of new messages or data.

## III. MACRO-ITERATION SEQUENCE

When studying asynchronous iterations, it is especially important to focus on the concept of macro-iteration sequence. Macro-iteration sequence is derived from the concept of R-chaotic sequence introduced by Miellou (see [14], p. 63) in the context of chaotic relaxation. In [15], a new stopping criterion that relies on macro-iteration sequence is proposed for linear perturbed asynchronous iterations. Macro-iterations sequence is also commonly used for studying the convergence of parallel or distributed asynchronous iterations [22].

We recall the definition of macro-iteration sequence.

Definition 2: Consider: $l(j) = \min_{h \in \{1,\ldots,n\}} \{l_h(j)\}$, then, the macro-iteration sequence $\{j_k\}_{k \in \mathbb{N}}$, is defined as follows:

$$j_0 = 0,$$

$$j_{k+1} = \min_j \left\{ \bigcup_{j_k \leq l(r) \leq r \leq j} S_r = \{1, \ldots, n\} \right\}.$$

The concept of macro-iteration sequence presents the interest to introduce a valuable sequence of macro-labels $\{j_k\}, k = 1, 2, \ldots$

It follows from Definition 2, that each update $x_i(j)$ at label $j$, with $j_{k+1} \leq j$, is guaranteed to use values $x_1(l_1(j)), \ldots, x_n(l_n(j))$ which satisfy:

$$j_k \leq l_i(j), 1 \leq i \leq n.$$

In the worst case, the values of the components of the iterate vector that are used during an updating phase correspond to labels located at the previous macro-iteration. This point is convenient for structuring the sequence of labels $\{j\}, j = 1, 2, \ldots$ in particular when studying the convergence of asynchronous iterations or for establishing stopping criteria (see [15] and [23]).

Since the theoretical work of Miellou, the concept of macro-iteration has been widely used in the literature in order to show the convergence of asynchronous iterations (with unbounded delays and possible out of order messages) for applications that range from numerical simulation and Markov systems to convex optimization. In the special case of numerical simulation, the reader is referred to in particular [22], where macro-iterations are used for convergence analysis as well as convergence detection purpose.

The General Convergence Theorem of Bertsekas for totally asynchronous iterative algorithms, i.e. asynchronous iterations with unbounded delays and out of order messages, also relies on the concept of macro-iteration sequence (see [3] and [11, p. 431]). The theorem which considers Cartesian product of sets and introduces level sets, i.e., boxes, is a powerful tool to prove convergence of asynchronous iterations for many applications. The basic idea is that from one macro-iteration to the next one, the sequence of iterate vectors, which starts from a level set (a box), enters the next box that is smaller and consequently progresses towards the solution of the problem, i.e., the fixed point that is situated at the intersection of all level sets.

The convergence of asynchronous iterations relies on mathematical properties of the associated operator $F$:
- monotonicity and continuity, e.g., see [4],
- or contraction, e.g. see [1].

With this regards, macro-iteration sequences have been widely used in the domain of convex optimization. There are many contributions in the literature on the convergence of distributed or parallel asynchronous iterative algorithms (with unbounded delays and out of order messages) for various optimization problems. In particular, the convergence of distributed asynchronous relaxation methods, i.e., distributed asynchronous iterative algorithms with unbounded delays and possible out of order messages, has been proposed for convex optimization problems in [6]. The convergence of asynchronous gradient algorithms (with unbounded delays, possible out of order messages and fixed step-size) is shown in [8], where we have considered convex separable network flow problems.

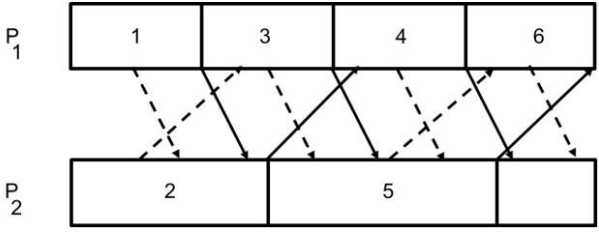

Fig. 2 Example of asynchronous iterative algorithm with flexible communication.

## IV. ASYNCHRONOUS ITERATIONS WITH FLEXIBLE COMMUNICATION

In [9], we have proposed a new class of parallel or distributed asynchronous iterative algorithms with flexible communication. The monotone convergence of this new class of asynchronous iteration is also studied in [9]. The new class of parallel or distributed iterative method extends asynchronous iterations. It does not necessarily make use of values of components on the iterate vector that are labelled by an iteration number such as $x_i(l_i(j))$, it can also use partial updates of the iterate vector in the computation.

The behavior of asynchronous iterations with flexible communication is shown in Figure 2 where communications of partials updates of the iterate vector are displayed as hatched lines with arrow.

As a consequence, asynchronous iterations with flexible communication are particularly well suited to the context of monotone convergence (updating phases can immediately take benefit of partial updates) and block iterative methods (partial results of computation can be used before the end of an iteration, i.e., before obtaining a new block update). This new class of asynchronous iterations is also well suited to modern communication operations such as one-sided MPI or direct memory access via put() or get() communication operations. Another interest of asynchronous iterations with flexible communication (which are sometimes called asynchronous iterations with order intervals, e.g., see [23]) is that they use approximate operators.

We now present the definition of asynchronous iterations with flexible communication [24] (see also [9] and [23]).

Definition 3: Let $F$ be an operator from $\mathbb{R}^n$ to $\mathbb{R}^n$. Let $G$ an operator from $\mathbb{R}^n$ to $\mathbb{R}^n$ such that for all $i \in \{1, ..., n\}$, $G_i(x)$ approximates $F_i(x)$. An asynchronous iteration with flexible communication $(G, x(0), \mathcal{S}, \mathcal{L})$ associated to operator $G$ and starting with iterate vector $x(0)$ is a sequence: $\{x(j)\}, j = 0, 1, ...,$ of vectors of $\mathbb{R}^n$ defined recursively by:

$$x_i(j) = \begin{cases} G_i(\tilde{x}_1(j), ..., \tilde{x}_n(j)) \text{ if } i \in \mathcal{S}_j, \\ x_i(j-1) \text{ if } i \notin \mathcal{S}_j, \end{cases} \quad (2)$$

where, $\mathcal{S}$ and $\mathcal{L}$ are subject to conditions a), b) and c) given in section II. Moreover, the values $\tilde{x}_1(j)$ of the components of the iterate vector (which correspond to exchanged data) must satisfy the following norm constraint:

$$\frac{\|\tilde{x}_i(j) - x_i^*\|_i}{u_i} \leq \|x(l(j)) - x^*\|_u, \forall i \in \{1, ..., n\}, \quad (3)$$

where $x(l(j))$ denotes the vector $\left(x_1(l_1(j)), ..., x_n(l_n(j))\right)$.

We note that the values $\tilde{x}_i(j)$ in (2) and (3) can either correspond to updates (which are represented by arrows in Figure 2) or partial updates (which are depicted by hatched arrows in the same figure).

Asynchronous iterations with flexible communication feature also unbounded delays and possible out of order messages since conditions a), b) and c) are satisfied. More details on asynchronous iterations with flexible communication can be found in [24] (see also [9] and [23]).

Asynchronous iterations with flexible communication are based on the use of operator $G$ which approximates the operator $F$. We note that the operators $G$ can be generated via an iterative process.

The monotone convergence of distributed asynchronous approximate gradient algorithms with flexible communication and fixed step-size was shown in [9] for a class of convex optimization problems, i.e., convex network flow problems.

The implementation of asynchronous approximate gradient algorithms with flexible communication was carried out on a Tnode parallel machine with distributed memory architecture, see [9]. Numerical results are also presented in [9]. The implementation on the Cray T3E supercomputer of asynchronous gradient type algorithms via the put() function of the SHMEM library of Cray is presented in [10], where the case of flexible communication is also considered. Computing results for convex optimization problems show the great efficiency of asynchronous gradient type algorithms. Flexible communication permits one to improve efficiency of asynchronous gradient algorithms.

The convergence of asynchronous modified Newton and Newton multi-splitting method with flexible communication was shown in [25] for convex network flow problems.

We underline that the convergence of asynchronous iterative algorithms with flexible communication has also been shown for contracting operators in [24].

Convex optimization problems are not the sole domain of application where distributed or parallel asynchronous iterative algorithms exhibit excellent efficiency. We have obtained excellent efficiency and scalability for applications related to numerical simulation. In particular, asynchronous iterative algorithms performing a huge amount of data exchanges for the solution of the obstacle problem have been carried out with success in real conditions on several supercomputers such as the IBM SP4 supercomputer (see [26], where several data exchange frequencies have been studied).

We have also considered distributed computing platforms such as Planetlab, whereby computations were carried out on several computing nodes scattered on different continents. In particular, computing experiments were performed on a network with nodes located in Australia and North America, see [27] and optimization as well as numerical simulation problems were solved thanks to the cooperation of these distant computing nodes. Finally, we have considered clusters and grid computing infrastructures such as GRID5000 (see [28]), whereby computations were distributed among several clusters situated in different cities in a multicore and multi network configuration, i.e., a configuration that combines Infiniband, Myrinet and fast Ethernet network for the same computing application. In all cases, asynchronous iterative algorithms have exhibited interesting efficiencies. In particular, efficiency and scalability of asynchronous iterations was better than the one of their synchronous counterparts. See also [29], for other computational tests on supercomputer architectures.

Recently, Mishchenko, Iutzeler and Malick have considered the solution of a class of convex optimization problems via distributed flexible asynchronous proximal gradient algorithms with unbounded delays [30]. The authors introduce the concept of epochs sequence: $\{k_m\}_{m\in\mathbb{N}}$ with

$$k_0 = 0,$$

and

$$k_{m+1} = \min_{k} \left\{ \begin{array}{c} \text{such that each machine} \\ \text{made at least two updates} \\ \text{on interval } \{k_m, k\} \end{array} \right\}$$

In particular, Mishchenko, Iutzeler and Malick claim that "in order to subsume delays, they develop a new epoch-based mathematical analysis, encompassing computation times and communication delays, to refocus the theory on algorithmics." They note also that "the main feature of the epoch sequence is that it automatically adapts to variations of behaviors of machines across time (such as one worker being slow at first that gets faster with time). The sequence then allows for intrinsic convergence analysis without any knowledge of the delays, as shown in the previous sections. This simple but powerful remark is one of the main technical contributions of the paper." They also add in the conclusion of their paper "these special features lead us to two key theoretical findings: i) an epoch-based analysis adapted to any kind of delays…"

We underline that the concept of epoch or meta-iteration of Mishchenko, Iutzeler and Malick is less general than the concept of macro-iteration sequence that was introduced a long time ago (see [15]). In particular, macro-iteration sequences account for possible out of order messages while epochs do not. In a macro-iteration, all components of the iterate vector are updated at least one time using available values of the components of the iterate vector that are associated to the previous macro-iteration.

Miellou [14] as well as Mishchenko, Iutzeler and Malick [30] have considered the case where for all $i \in \{1, \dots, n\}$, the functions relative to delayed iterates $l_i: \mathbb{N} \to \mathbb{N}$, such that $l_i(j) = j - d_i(j)$, are monotone increasing. Miellou has considered bounded communication delays (see [14], p. 63) and unbounded communication delays (see [15]). Mishchenko, Iutzeler and Malick have considered unbounded communication delays in [30]. The paper [30] deals with distributed asynchronous gradient algorithms with unbounded delays for a special class of convex optimization problems with a smooth separable function and a non-smooth function. In subsection 3.1, the authors say: "To the best of our knowledge, all papers on asynchronous distributed methods… assume that delays are uniformly upper bounded by a constant." Mishchenko, Iutzeler and Malick do not quote the work by Bertsekas and El Baz on distributed asynchronous relaxation method for convex network flow problems (see [6]), nor the papers of El Baz on distributed asynchronous gradient type methods for convex optimization problems, e.g., [8] as well as distributed asynchronous iterations with flexible communication for convex optimization problems, e.g., see [9] and [25] and numerical simulation problems, e.g., see [15], [23] and [24]. In particular, we note that references [6] to [9] and [23] to [25] deal with asynchronous iterations with unbounded delays. Moreover, references [9] and [23] to [25] present flexible communication and mathematical operators which are approximations (generated via iterative processes) of fixed-point operators (see also [31]).

V. CONVEX OPTIMIZATION AND MACHINE LEARNING APPLICATIONS

We consider the problem:

$$\min_{x\in\mathbb{R}^n} f(x) + g(x), \qquad (4)$$

where $f(x)$ is a separable, $L$-smooth (with $L \geq 0$), $\mu$-strongly convex function (with $\mu > 0$) and $g$ is a lower semi-continuous non-smooth convex function (see [30] and [32]).

Problem (4) describes a variety of machine learning problems as well as signal processing applications. For example, one can have : $\{(y_h, z_h)\}_{h=1}^m$, a set of $m$ training samples, each consisting of input $y_h$ and target $z_h$. The objective is to learn the parameters $x$ of the model $p(y, x)$ so that $p(y_h, x)$ matches the target $z_h, h = 1,2, \dots m$. Some loss function $h$ gives a measure on how well a prediction $p(y_h, x)$ matches the target $z_h$. We can have: $f_h(x) = h(p(y_h, x), z_h)$. We use the regularization function $g(x)$ in order to avoid over-fitting the training data.

Let $\gamma \in \left(0, \frac{2}{\mu+L}\right]$ be the fixed gradient step-size associated to the function $f$.

Let the proximal operator of the non-smooth convex function g be defined as follows:

$$prox_{\gamma,g}(x) = \arg\min_{v} \left\{ g(v) + \frac{1}{2\gamma} \|v - x\|^2 \right\}.$$

We define the approximate gradient-type operator G with fixed step-size.

Definition 4: The approximate gradient-type operator $G: \mathbb{R}^n \to \mathbb{R}^n$ associated to the convex optimization problem (4) is such that for all $i \in \{1, \dots, n\}$, we have:

$$x_i(j) = G_i\left(x_1(l_1(j)), \ldots, x_n(l_n(j))\right)$$
$$= prox_{\gamma,g}\left(x_1(l_1(j)), \ldots, x_n(l_n(j))\right)$$
$$- \gamma \frac{\delta f(prox_{\gamma,g}(x_1(l_1(j)), \ldots, x_n(l_n(j))))}{\delta x_i}.$$

Theorem 1: The asynchronous iteration with flexible communication $(G, x(0), \mathcal{S}, \mathcal{L})$ associated to the gradient-type operator $G$ satisfies for all $j_k \leq j$:

$$\|x(j) - x^*\|^2 \leq (1-\rho)^k \max_{1 \leq i \leq n} \|x_i(0) - x^*\|^2, \quad (5)$$

where

$$\rho = \gamma \cdot \mu,$$

and $\{j_k\}_{k \in \mathbb{N}}$ is the macro-iteration sequence introduced in Definition 2. Moreover, the asynchronous iteration with flexible communication $(G, x(0), \mathcal{S}, \mathcal{L})$ converges to the solution $x^*$ of the problem (4).

Proof: inequality (5) follows from (3), Definition 4, and Lemma 3.1 in [22]. The convergence of asynchronous iterations with flexible communication follows from (5) and Theorem 2 in [24].

Remark 1: we note that the convergence result for asynchronous iterations with flexible communication relies on the contraction property of the gradient-type operator.

Remark 2: for simplicity of presentation, we have considered in Definition 4, an approximate gradient-type operator which performs only one gradient-type iteration with fixed step-size. We can also consider approximate gradient-type operators which perform several gradient-type iterations with fixed step-size and we can show similarly the convergence of asynchronous iterations with flexible communications for those approximate operators, e.g., see Lemma 1 in [24].

Remark 3: parallel or distributed asynchronous iterative algorithms with flexible communication are particularly interesting in the context of machine learning applications whereby training is cumbersome and models are complex and large. The possibility to distribute computation on many computing nodes that belong to hybrid architectures or distributed heterogeneous systems and to scale up thanks to asynchronous computation is especially attractive.

Finally, we refer to [31] for a study relative to training many neural networks in parallel with Graphic Processing Units (GPU) via Back-Propagation. We recall that GPUs are massively parallel computing accelerators.

## VI. Conclusions

In this paper we have presented and commented on important concepts related to the study of parallel or distributed asynchronous iterative algorithms. In particular, we have considered the concepts of unbounded delays, out of order messages, macro-iteration sequence and flexible communication. We have presented the advantages of parallel or distributed asynchronous iterations. Finally, we have given a new result of convergence for parallel or distributed asynchronous iterations with flexible communication, for convex optimization and machine learning.

In future work, we plan to carry out on the Grid5000 platform asynchronous iterative algorithms with flexible communication presented in this paper for machine learning. We shall use GRIDHPC, a decentralized environment for High Performance Computing that we have designed and developed in a multicore and multi network configuration.

Acknowledgment

Part of this study has been made possible via funding of Centre International de Mathématiques et d'Informatique de Toulouse: CIMI-ANR-11-LABX-0040 - LABX-2011.